\documentclass[10pt]{article}

\usepackage{amsmath}
\usepackage{array}
\usepackage{appendix}
\usepackage{graphicx}
\usepackage{amsfonts}
\usepackage{amssymb}
\usepackage{mathrsfs}
\usepackage{yfonts}
\usepackage{euscript}
\usepackage{upgreek}
\usepackage{slantsc}
\usepackage{calligra}
\usepackage[T1]{fontenc}
\usepackage{epsf}
\usepackage{latexsym}

\usepackage{tipa}

\def\be{\begin{equation}}
\def\ee{\end{equation}}
\def\beq{\begin{equation}}
\def\eeq{\end{equation}}
\def\bea{\begin{eqnarray}}
\def\eea{\end{eqnarray}}

\def\!{\hspace{-1.6667em}}

\def\me{\mbox{e}}

\def\mh{\mbox{h}}

\def\ml{\mbox{l}}   
   
\def\mn{\mbox{n}}

\def\fQ{\mbox{\sffamily Q}}

\def\bn{\mbox{\bf n}}

\def\bn{\mbox{{\bf n}}}
\def\bq{\mbox{{\bf q}}}

\def\sa{\mbox{\scriptsize a}}

\def\scc{\mbox{\scriptsize c}}

\def\se{\mbox{\scriptsize e}}
\def\sf{\mbox{\scriptsize f}}
 
\def\sh{\mbox{\scriptsize h}} 
\def\si{\mbox{\scriptsize i}}

\def\sll{\mbox{\scriptsize l}}  
\def\sm{\mbox{\scriptsize m}}
\def\sn{\mbox{\scriptsize n}} 
\def\so{\mbox{\scriptsize o}}

\def\sr{\mbox{\scriptsize r}}
\def\sss{\mbox{\scriptsize s}}  
\def\st{\mbox{\scriptsize t}}

\def\sB{\mbox{\scriptsize B}}

\def\sG{\mbox{\scriptsize G}}

\def\sJ{\mbox{\scriptsize J}}
\def\sK{\mbox{\scriptsize K}}

\def\sR{\mbox{\scriptsize R}}

\def\sW{\mbox{\scriptsize W}}

\def\sfA{\mbox{\sffamily{\scriptsize A}}}
\def\sfB{\mbox{\sffamily{\scriptsize B}}}
\def\sfC{\mbox{\sffamily{\scriptsize C}}}

\def\sbq{\mbox{{\bf \scriptsize q}}}

\def\ta{\mbox{\tiny a}}

\def\te{\mbox{\tiny e}}

\def\th{\mbox{\tiny h}}
\def\ti{\mbox{\tiny i}}

\def\tl{\mbox{\tiny l}}

\def\tn{\mbox{\tiny n}}
\def\to{\mbox{\tiny o}}

\def\tr{\mbox{\tiny r}}

\def\ttt{\mbox{\tiny t}}   

\def\tG{\mbox{\tiny G}}

\def\tR{\mbox{\tiny R}}

\def\tfC{\mbox{\sffamily{\tiny C}}}

\def\lt{\mbox{\Large $t$}}

\def\K{Kucha\v{r} }

\def\pa{\partial}
\def\d{\textrm{d}}

\def\Last{\mbox{\Large$\ast$}}                
\def\5Star{\mbox{\Large$\star$}}              
\def\sumi3{\sum\mbox{}_{\mbox{}_{\mbox{\scriptsize $i$=1}}}^3}

\def\sumj3{\sum\mbox{}_{\mbox{}_{\mbox{\scriptsize $j$=1}}}^3}
\def\sumk3{\sum\mbox{}_{\mbox{}_{\mbox{\scriptsize $k$=1}}}^3}

\begin{document}

\begin{titlepage}

\begin{center}
\Large{\bf MINISUPERSPACE MODEL OF MACHIAN RESOLUTION} 
\vspace{0.05in} 
{\bf OF PROBLEM OF TIME. I. ISOTROPIC CASE} \normalsize

\vspace{.1in}

{\large \bf Edward Anderson}

\vspace{.1in}

{\large {\em DAMTP Cambridge \normalsize}}, ea212@cam.ac.uk

\end{center}

\begin{abstract}

A local resolution to the Problem of Time that is Machian and was previously demonstrated for relational mechanics models is here shown to work for a 
more widely studied quantum cosmological model. 
I.e., closed isotropic minisuperspace GR with minimally-coupled scalar field matter.  
This resolution uses work firstly along the lines of Barbour's at the classical level. 
Secondly, it uses a Machianized version of the Semiclassical Approach to Quantum Cosmology (the resolution given is not more than semiclassical).  
Finally, it uses a Machianized version of a combined Semiclassical Histories Timeless Records scheme along the lines of Halliwell's work.  
This program's goal is the treatment of inhomogeneous perturbations about the present paper's model. 
This draws both from this paper's minisuperspace work and from qualitative parallels with relational particle mechanics, 
since both have nontrivial notions of inhomogeneity/structure (clumping) as well as nontrivial linear constraints.  

\end{abstract}


\section{Introduction}\label{Intro}

In this paper, I demonstrate a local resolution to the Problem of Time (PoT) \cite{Kuchar92I93, APOTAPOT2, FileR} that is Machian in character 
\cite{BB82, B94I, RWR, ARel, ARel2, FileR}. 
See Sec 2 for explanations of the facets that constitute the PoT facets, strategies resolving them, 
and for the precise meaning given here to the terms `Machian' and `a local resolution' 
To be clear, this resolution applies in semiclassical regimes. 
Whilst this is a theoretical limitation, the semiclassical quantum cosmology regime is of more immediate observational relevance than more fully quantum regimes.  

\mbox{ }  

\noindent I previously demonstrated such a Machian local resolution of the PoT for relational particle models (RPM's) \cite{AHall, ACos2, FileR, CapeTown12, QuadI, QuadIII}.
The minisuperspace \cite{Magic, Ryan, IB75, HH83etc} (alias homogeneous GR) case is in many ways more straightforward (no nontrivial linear constraints).   
It is relevant both as an approximate model for our universe and as the zeroth order of more accurate such models along the lines of Halliwell--Hawking \cite{HallHaw}. 
The latter are the next step in the present program; the extent to which our local resolution to the PoT extends to this is the subject of work in progress (see the Conclusion for more).
The Halliwell--Hawking model is of inhomogeneous perturbations about homogeneous cosmology. 
It is already a realistic model for the possible quantum-cosmological origin of large-scale structures such as galaxies and CMB hot-spots 
(via amplification by some inflationary mechanism \cite{inflation-13}).
This model is like RPM's in possessing nontrivial linear constraints, albeit now with further nontrivial diffeomorphism connotations.
Additionally in both cases the linear constraints are algebraically solvable \cite{ASand}. 
Thus working with a model along the lines of Halliwell--Hawking's \cite{ASand, A-HallHaw-1} 
is jointly the natural successor of the present paper's minisuperspace work and previous papers' RPM work.

\mbox{ }

\noindent I am drawn to working with specifically the spatially-$\mathbb{S}^3$ Friedmann--Lema\^{i}tre--Robertson--Walker (FLRW) cosmology by the following.
i) A combination of Machian considerations and simplicity.  
ii) I am moving toward working with a model along the lines of Halliwell--Hawking's more realistically cosmological model.
Additionally I am drawn to include scalar field matter in this work by the following.
a) Standard practise in Cosmology.
b) The specifics of the Halliwell--Hawking model. 
c) That models with one degree of freedom are relationally trivial (see also Sec 2 for what I mean by `relational' in this paper.)
A cosmological constant term is needed to support the spatially-$\mathbb{S}^3$ FLRW cosmology with scalar field matter in the case in which matter effects are presumed small 
\cite{Rindler}.
There are then two different sets of modelling assumptions to consider. 
1) The matter physics is light and fast (l) as compared to the gravitational physics being heavy and slow (h: not to be confused with the spatial 3-metric $h$). 
2) Both are h and only further entities introduced in further papers (anisotropy in \cite{A-MSS-2} or inhomogeneity in \cite{ASand, A-HallHaw-1}) are l.  
This use of h and l terminology is standard in semiclassical approaches, as detailed in Sec 3.  
Secs 2 and 3 explain PoT facets alongside their Machian resolution, using the isotropic minisuperspace model as an example in each case.  
With justification of the semiclassical approach's crucial WKB ansatz and approximation being tenuous, Sec 3 considers the combined Machian Semiclassical--Histories--Records scheme.
This rests on I) a small expansion upon Anastopoulos and Savvidou's work \cite{AS05} on Histories Theory for isotropic minisuperspace with scalar field. 
II) Halliwell's work \cite{H03, H09H11} on not explicitly Machian versions of such combinations.  
I conclude in Sec 4.  
A further paper will contain detailed perturbative equations both for this paper's case and for the more complicated and generic case of Bianchi IX Minisuperspace too \cite{A-MSS-2}.  

\end{titlepage} 

\section{Machian local resolution of classical Problem of Time}

\subsection{Temporal Relationalism}\label{TR}

\noindent Temporal Relationalism is the Leibnizian idea that `there is no time for the universe as a whole' \cite{B94I, EOT, FileR}.  
This is desirable as regards background independence and the modelling of closed universes.
The well-known quantum-level Frozen Formalism Problem facet of the Problem of Time 
-- that GR's quantum wave equation is timeless -- then turns out to be a {\sl consequence} of Temporal Relationalism, as we shall see below.  
Moreover, the rest of this SSec makes it clear that Temporal Relationalism is already present at the classical level; 
this rather less well-known point is in fact the essential content of Barbour's \cite{B94I}.   

\noindent Note that in the Temporal Relationalism idea, no details of what is meant by `universe' are required other than it being a whole system rather than a subsystem. 
(Thus how our conception of universe has changed since Leibniz's day due to GR and modern observational cosmology does not affect use of this idea.) 

\noindent The precise meaning of Temporal Relationalism comes, rather, from adopting the following sharp mathematical implementation. 

\noindent 1) The action is not to contain any extraneous time or time-like variables. 

\noindent 2) The action is to be reparametrization-invariant.  
The usual role of time in actions has been replaced by a mere label time that is physically meaningless; this is the parameter in question.
It then makes more sense for primary-level conceptualization to not be centred about physically meaningless entities. 
So instead of writing the action in terms of a meaningless parameter, one writes it down without resorting to making a parametrization, i.e. a parametrization-irrelevant action. 
Finally, it is still better not to name such an action after an absent irrelevant entity, and one can do so because the actions in question happen to be, dually, 
geometrical arc elements.

\noindent Thus what I actually postulate is a geometrical arc element action (of the Jacobi--Synge type as detailed in the next paragraph). 
This then implies, dually, parametrization-irrelevance, which in turn implies reparametrization-invariance upon introduction of a parameter.
This parameter is then identified as a label-time that has supplanted the primary-level role played by absolute time in non-relational physics,  
so that the stated postulate indeed succeeds in making contact with a change in how time is modelled in physical theories.  

\mbox{ }

\noindent Example 1) the Jacobi action for the mechanics of configurations $Q^{\sfA}$ (e.g. positions of $N$ particles) is
$$
S  = \sqrt{2}\int \d \widetilde{s} 
  := \sqrt{2}\int \d s\sqrt{E - V(Q^{\sfC})} 
  := \sqrt{2}\int \sqrt{M_{\sfA\sfB}\d Q^{\sfA} \d Q^{\sfB}} \sqrt{E - V(Q^{\sfC})}
$$
\beq
   =       2 \int \d\lambda \sqrt{\frac{M_{\sfA\sfB}}{2} \frac{\d Q^{\sfA}}{\d \lambda}\frac{\d Q^{\sfB}}{\d\lambda}} \sqrt{E - V(Q^{\sfC})}
  :=       2 \int \d\lambda \sqrt{T} \sqrt{E - V(Q^{\sfC})} \mbox{ } . 
\label{S-Jac}  
\eeq
The first form is in terms of the physical line element $\d \widetilde{s}$, and obviously constitutes a {\it geodesic principle}.
The second form is in terms of the kinetic line element $\d s$, which is conformally-related to the preceding by a conformal factor $\sqrt{E - V}$. 
Thus this is a geodesic principle up to a conformal factor, which is termed a parageodesic principle.
The third form defines $\d s$; $M_{\sfA\sfB}$ is the kinetic metric.
The fourth form then demonstrate the insertion of a parameter by which the first form is parametrization-irrelevant and the second form is reparametrization-invariant.
The fifth form recasts the arc element as a kinetic energy term whose velocities\footnote{The third form's changes in configuration $\d Q^{\tfC}$ 
are held to be more primary than these velocities through not involving the label time.}
are built out of label-time derivatives, producing a form more familiar to some readers.
Thus it is clear that the Jacobi action corresponds to theories whose kinetic terms are purely quadratic in their velocities.

\mbox{ }  

\noindent Example 2) Synge subsequently considered more general geometrical arc elements that are not subjected to this purely quadratic restriction \cite{Lanczos}.

\mbox{ } 

\noindent Example 3) Full GR with a minimally-coupled scalar field then has a similar relational action (along the lines of \cite{BSW, RWR, FileR})\footnote{In this paper, 
for further Machian reasons independent from the present SSec's discussion of Mach's Principle, the spatial topological manifold $\Sigma$ is taken to be compact without boundary.
This is to avoid undue influence of boundary or asymptotic physics, and is a criterion also argued for by Einstein.
It is also a simpler case to handle mathematically.
Moreover, the simplest particular such $\Sigma$ is the $\mathbb{S}^3$ adopted by Halliwell--Hawking and hence also in the present paper.}
\beq
S^{\sG\sR}_{\sr\se\sll\sa\st\si\so\sn\sa\sll} = \int \int_{\Sigma} \d^3x \sqrt{h} \d s \sqrt{\mbox{Ric}(x; h] - 2\Lambda - |\pa\phi|^2 - V(\phi)} 
\mbox{ } , \mbox{ }  \mbox{ }  
\d s = ||\d_{F}(h, \phi)||_{\mbox{\boldmath\scriptsize $\cal M$}(h)}
\mbox{ } .  
\label{S-BSW-Type}
\eeq
This has blockwise configuration space metric ${\mbox{\boldmath $\cal M$}(h)} := \mbox{\Huge{(}} \stackrel{\mbox{\normalsize $M(h)$ \,\, 0}}{\mbox{\normalsize \, \, 0  \, \, \, 1}} 
\mbox{\Huge{)}}$ and $M(h)$ the GR configuration space metric (alias inverse DeWitt supermetric).
$||\d_{F}(h, \phi)||_{\mbox{\boldmath\scriptsize $\cal M$}}$ is the corresponding quadratic form.\footnote{I use round brackets for function dependence, 
square brackets for functional dependence and $( \mbox{ } ; \mbox{ } ]$ for mixed function dependence (before the semicolon) and functional dependence (after the semicolon).
$h_{ab}$ is the spatial 3-metric, with determinant $h$, Ricci scalar Ric$(x; h]$ and GR configuration space metric $M^{abcd} := h^{ac}h^{bd} - h^{ab}h^{cd}$.
The present paper's $\mathbb{S}^3$ closed-universe case requires the cosmological constant term $\Lambda$ for nontrivial solutions to exist.
$\phi$ is a minimally-coupled scalar field. See Sec 2.5 for the meaning of the $F$ subscript.}  

\mbox{ } 

\noindent Note 1) Using (\ref{S-Jac})       produces the same mathematics as arises from the Euler--Lagrange formulation of mechanics, 
              and using (\ref{S-BSW-Type})  produces the same mathematics as arises from the Arnowitt--Deser--Misner (ADM) dynamical formulation of GR. 
These equivalences are exposited in e.g. \cite{Lanczos, BSW, FEPI, FileR}.  
What the more familiar Euler--Lagrange formulation fails to do is be free ab initio from the extraneous Newtonian time.
The more conventional ADM action, on the other hand, fails to be free of the example par excellence of extraneous time-like variable: the lapse function $\alpha$.
Thus these more familiar actions fail to implement criterion 1) of Temporal Relationalism.  

\noindent Note 2) The title of \cite{BSW} -- `3-geometry as carrier of information about time' -- refers to 
the presence of the aforementioned duality between time and configuration space geometry, which for GR are spatial geometries.
Generalizing to arbitrary classical continuum theories and slightly expanded to highlight its temporally Machian character, 
this title becomes `configuration and change of configuration as carrier of information about time'.

\mbox{ } 

\noindent Example 4) This paper's isotropic minisuperspace is both a simplification of (\ref{S-BSW-Type}) and of a form similar to (\ref{S-Jac}): 
\beq
S_{\sr\se\sll\sa\st\si\so\sn\sa\sll}^{\mbox{\scriptsize isotropic-mss}} = \int \d s \, \mbox{exp}(3\Omega)\sqrt{\mbox{exp}(-2\Omega) - V(\phi) - 2\Lambda} 
\mbox{ } , \mbox{ }  \mbox{ }  
\d s = \sqrt{- \d\Omega^2 + \d\phi^2} \mbox{ } . 
\eeq
Here, the {\it Misner variable} $\Omega := \mbox{ln} a$ for $a$ the usual scale factor.  
The symmetry-reduced GR configuration space is, mathematically, just 2-$d$ Minkowski space $\mathbb{M}^2$ equipped with its standard indefinite flat metric.  

\mbox{ }

\noindent Dirac's argument \cite{Dirac} (originally made for reparametrization-invariant actions) then guarantees that such temporally relational actions must lead to primary constraints, 
i.e. relations between the momenta that arise purely from the form of the geometrical arc element.\footnote{One says `of the Lagrangian' in the reparametrization-invariant formulation, 
but parametrization-irrelevant and arc element conceptualizations use Jacobi--Synge arc elements instead of Lagrangians: $\d s$ in place of $L \, \d\lambda$ as the integrand of the action.}
%
($h, \phi$) denotes $h_{ij}$ and $\phi$: the the set of configuration variables.  
For this paper's case (and all other purely quadratic kinetic term theories), the square-root form of the action gives a GR Hamiltonian constraint ${\cal H}$,\footnote{In the ADM
formulation, ${\cal H}$ is well-known to arise from variation with respect to the lapse.  
In relational product-type actions like (\ref{S-Jac}), there is no longer any lapse. 
However the ensuing mystery of how to arrive at ${\cal H}$ in this case is straightforwardly mopped up by Dirac's observation that such an action guarantees a primary constraint.}
which, for the present paper's closed isotropic minisuperspace case, is 
\beq
{\cal H} := \mbox{exp}(-3\Omega)\{-\pi_{\Omega}^2 + \pi_{\phi}^2\} + \mbox{exp}(3\Omega)\{V(\phi) + 2\Lambda - \mbox{exp}(-2\Omega)\} = 0 \mbox{ } . 
\eeq 
Having a constraint purely quadratic in the momenta such as the GR Hamiltonian constraint indeed guarantees\footnote{In the case of the relativistic particle, 
a Klein--Gordon type wave equation arises instead. 
However, despite GR's redundant configuration space of spatial 3-metrics Riem($\Sigma$) being of indefinite signature, a Klein--Gordon interpretation cannot be pinned upon quantum GR. 
This is due to a lack of Killing vector structures that additionally manage to be compatible with GR's potential \cite{Kuchar81}.
Thus my `guarantees' precludes this alternative.}
the appearance of the quantum-level Frozen Formalism Problem.

\noindent Moreover, the above timelessness at the primary level can in the classical case be resolved at a secondary, emergent level by {\it Mach's Time Principle}: 
`time is to be abstracted from change'. 
Three distinct specifications of which involve `any change' (Rovelli \cite{Rfqxi}), `all change' (Barbour \cite{Bfqxi}) and my 
`sufficient totality of locally relevant change' (STLRC) \cite{ARel2}.  
\noindent This emergent time represents a local generalization of the astronomers' ephemeris time \cite{Clemence}; this is particularly manifest in the case of mechanics.
\noindent To fulfil the true content of the STLRC approach, all change is given opportunity to contribute to the timestandard. 
However only changes that do so in practise to within the desired accuracy are actually kept.\footnote{As argued in \cite{ARel2},  
this necessitates a curious indirect procedure in making such an approximation.  
I.e., one can not simply compare the sizes of the various energy terms, but must rather \cite{ARel2} assess this at the level of the resulting force terms upon variation.}  

\noindent All three of these specifications have some sense in which they are `democratic'.     
The senses in which `any change' and `all change' are democratic are clear. 
For STLRC, the democracy is in all change having {\sl the opportunity} to contribute. 
However, then only those changes whose contributions lead to effects above the desired accuracy are actually kept in practise. 
Additionally, only `all change' and STLRC take into consideration that `some clocks/timestandards are better than others' is an essential part of accurate timekeeping. 
(\cite{Bfqxi} explains how sidereal time gives more accurate predictions than (apparent) solar time as an example of this.)  
Finally, only `any change' and STLRC are operationally realizable.  
[`All change' fails in this regard since some of the universe's contents are inaccurately or completely unknown, so one can not include `all change' in accurate practical calculations.]
Thus overall, STLRC wins out.

\noindent Emergent Jacobi time is an implementation of Mach's Time Principle.  
At first sight it is an `all change' implementation, but, in practice it is a STLRC one 
(for operational reasons: not all changes in the universe are kept in practise in forming the expression for $t^{\se\sm(\sJ)}$).
For isotropic minisuperspace, this emergent time takes the form 
\beq
\lt^{\se\sm(\sJ)} = \int \d s\left/\sqrt{\mbox{exp}(-2\Omega) - V(\phi) - 2\Lambda} \right. \mbox{ } . 
\label{plain-tem}
\eeq
The oversized notation $\mbox{\Large $t$}^{\se\sm(\sJ)} = t^{\se\sm(\sJ)} - t^{\se\sm(\sJ)}(0)$ is used to incorporate selection of `calendar year zero', $t^{\se\sm(\sJ)}(0)$. 
From inspecting how this timestandard features in the equations of motion of the system, it represents a relational recovery of the well-known notion of cosmic time.  
Thus the current Machian time program specifically selects cosmic time as the timestandard to use in classical Cosmology.  
Moreover it provides a formula for it which, at least in detail, has further dependence on anisotropies and inhomogeneities. 
[At this point in the argument, the Mechanics counterpart amounts to a relational recovery of Newtonian absolute time, whilst in the more general GR context picks up the local proper time.]

\subsection{Further detail of classical Machian PoT resolution: h--l split}

\noindent We make an h--l split is between heavy degrees of freedom and light degrees of freedom.
This is a classical parallel of 1) the procedure in Molecular Physics by which one solves for the electronic structure under the approximation that the much heavier nuclei stay fixed. 
2) A technically similar approximation procedure from Semiclassical Quantum Cosmology.  
The new idea is that the h degrees of freedom provide an approximate timestandard with respect to which the l degrees of freedom evolve.

\subsection{The h-approximation is not Machian}

\noindent Here the relational action is 
\beq
S_{\sr\se\sll\sa\st\si\so\sn\sa\sll}^{\mbox{\scriptsize isotropic-mss\,(h = $\Omega$)}} = \int \mbox{exp}(3\Omega)\d\Omega\sqrt{2\Lambda - \mbox{exp}(-2\Omega)} \mbox{ } .
\eeq
We have used here the approximations $\d\Omega^2 >> \d\phi^2$ and $|2\Lambda - \mbox{exp}(-2\Omega)| =: |W(\Omega)| >> |V(\phi)|$. 
I.e. the scale contributions to the change/kinetic term and to the potential term dominate over the matter scalar field contributions to each of these respectively.
The Hamiltonian constraint is then
\beq
{\cal H}^{\sh} := - \mbox{exp}(-3\Omega) \pi_{\Omega}^2 + \mbox{exp}(3\Omega)\{2\Lambda - \mbox{exp}(-2\Omega)\} = 0  \mbox{ } . 
\eeq
The emergent Jacobi time is
\beq
\lt^{\se\sm(\sJ)}_{(0)} = \int \d\Omega/\sqrt{2\Lambda - \mbox{exp}(-2\Omega)} = 
{2\Lambda}^{-1/2}\,\mbox{artanh}\big( \sqrt{\{2\Lambda - \mbox{exp}(-2\Omega)\}/2\Lambda}\big) \mbox{ } .
\eeq
However, pure-h expressions of this general form --- $\lt^{\se\sm(\sJ)} = {\cal F}[\mh, \d \mh]$ ---  
are unsatisfactory from a Machian perspective since they do not give l-change an opportunity to contribute.

\subsection{Including further correction terms does render the scheme Machian}

This deficiency is to be resolved by treating them as zeroth-order approximations in an expansion involving the l-physics too.
Expanding (\ref{plain-tem}), one obtains a novel expression of the bona fide Machian form $\lt^{\se\sm(\sJ)}_{(1)} = {\cal F}[\mh, \ml, \d \mh, \d \ml]$: 
\beq
\lt^{\se\sm(\sJ)}_{(1)} = \lt^{\se\sm(\sJ)}_{(0)} + \frac{1}{2}
\left\{
V(\phi)\int\frac{\d \Omega}{W_{\Omega}^{3/2}} -  
\int\frac{\d\Omega}{\sqrt{W_{\Omega}}} 
\left\{
\frac{\d \phi }{\d \Omega}
\right\}^2 
\right\} 
+ O\left(\left\{\frac{\d \phi}{\d \Omega}\right\}^4\right) + O\left(\left\{\frac{V_{\phi}}{W_{\Omega}}\right\}^2\right) \mbox{ } .
\label{Cl-Expansion}
\eeq
Here the leading corrections are, firstly a $\phi$-potential term, the integral in the expression of which comes out as 
\beq
-1/2\Lambda\sqrt{2\Lambda - \mbox{exp}(-2\Omega)}  +   \{1/2\} \mbox{artanh}(\sqrt{\{2\Lambda - \mbox{exp}(-2\Omega)\}/{2\Lambda}})/\sqrt{2\Lambda^3} \mbox{ } , 
\eeq
Secondly, there is also an l-change term that renders the scheme Machian in the above sense.  

\noindent A more full analysis involves expanding h, l, $t^{\se\sm(\sJ)}$ as functions of an n-tuple of small `$\epsilon$' quantities.
See \cite{A-MSS-2} for this in full (or \cite{ACos2} for the already-available RPM counterpart), 
but in outline these small quantities are l-to-h kinetic term ratios and potential term ratios, interaction term to h ratio and ratios of derivatives of such terms.  
A key feature here is that what is conventionally an `independent variable' $t$, is here a $t^{\se\sm(\sJ)}$ that is a priori a {\sl highly-dependent} variable.  
As such it is clear that this itself is to be subjected to perturbations, whereas the conventional `independent variable' use of $t$ itself is not.  
See \cite{A-MSS-2} for details of the equations that this produces in this minisuperspace case.  

\mbox{ }

\noindent By the Problem of Time, however, I do not merely mean the Frozen Formalism Problem, but rather the totality of QM--GR incompatibilities for which \K and Isham 
list the Frozen Formalism Problem as but one of eight facets.  
The remaining facets are outlined below.

\subsection{Configurational Relationalism}

%
In the context of GR, the second PoT facet is most usually termed the Thin Sandwich Problem  \cite{BSW, TSP, ASand}.   
The Sandwich name is specific to GR as geometrodynamics [see Fig 1 a) and b)].
The Thin Sandwich Problem consists of solving the Lagrangian variables ($Q^{\sfC}$, $\dot{Q}^{\sfC}$) form of the GR momentum constraint for the 3-diffeomorphism auxiliary variable 
(usually the shift).  
However, in the case of an arbitrary physical theory, the Thin Sandwich Problem is generalized by Barbour's Best Matching Problem \cite{BB82, FileR} [Fig 1 c) and d)]. 
This consists of solving the Lagrangian variables form of those of the theory's first-class constraints that are linear in the momenta for the corresponding auxiliary variables.
Such theories have a group $G$ of continuous motions that are taken to be physically irrelevant (this is what is meant by Configurational Relationalism). 
In Best Matching, then, the classical action is built to contains auxiliary variables corresponding to $G$. 
These are then to be removed by 
i) solving the constraints that arise from variation with respect to the $G$-auxiliaries (which are the aforementioned linear constraints) for the $G$-auxiliaries themselves. 
ii) One then substitutes these solutions back into the action itself to obtain an action that is $G$-invariant directly (i.e. without involving any $G$-auxiliaries) .

{            \begin{figure}[ht]
\centering
\includegraphics[width=1.0\textwidth]{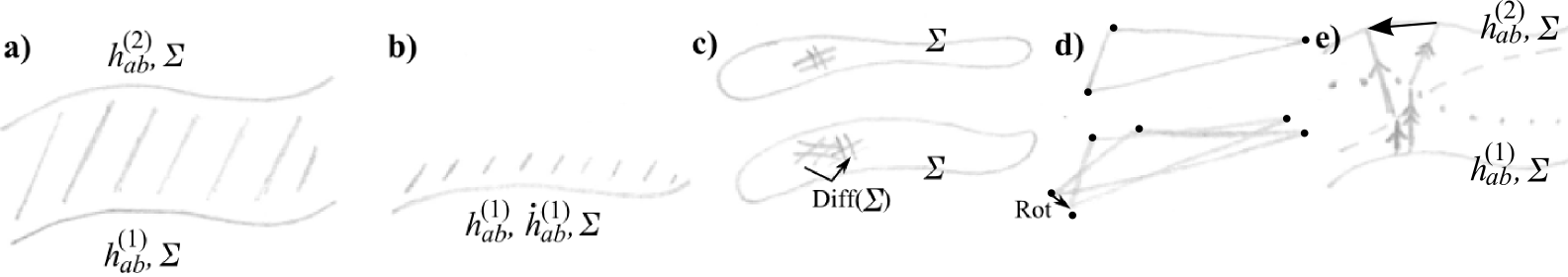}
\caption[Text der im Bilderverzeichnis auftaucht]{\footnotesize{a) Wheeler's thick sandwich involves solving the data on the bounding bread-slices for the spacetime `filling' (shaded). 
However, this failed to be well-posed. 
b) It was succeeded by Wheeler's Thin Sandwich \cite{BSW, TSP}: the infinitesimal limit of the previous, 
in which one solves metric and velocity of the metric data in order to construct a localized slab of spacetime (shaded).
c) The Thin Sandwich can then be re-interpreted in terms of Best Matching Riem($\Sigma$) with respect to Diff($\Sigma$). 
This amounts to the depicted comparison of two 3-metric configurations, 
by keeping one fixed and diffeomorphism-shuffling the other around until it is brought into maximum congruence with the fixed configuration.
d) This Best Matching description then applies to any other theory with redundant configurations, as exemplified by Barbour's keeping one triangle fixed 
and rotating the other around  around until it is brought into maximum congruence with the fixed triangle.  
e) Foliation Dependence in GR is not a problem because evolving from $\Sigma, h_{ij}^{(1)}$ via the dashed hypersurface to $\Sigma, h_{ij}^{(2)}$ coincides on $\Sigma, h_{ij}^{(2)}$ 
with the outcome of evolving it rather via the dotted hypersurface.
This result -- Refoliation Invariance -- holds because the two evolutions each correspond to applying the Hamiltonian constraint twice in different orders 
(grey arrow and then grey double arrow versus grey double arrow and then grey arrow). 
But that difference is therefore the commutator of two Hamiltonian constraints, and Dirac \cite{Dirac} established that this is just a momentum constraint term. 
This corresponds just to a diffeomorphism internal to $\Sigma, h_{ij}^{(2)}$ itself (the black arrow), hence establishing the result \cite{T73}.} }
\label{Facet-Intro} \end{figure}          }

\noindent Note that to render Best Matching compatible with Temporal Relationalism\footnote{See \cite{AM13} 
for an outline of how to render the entirety of the Principles of Dynamics compatible with Temporal Relationalism.}
\cite{ABFO, FEPI, FileR}, multiplier coordinate auxiliaries such as the shift to be replaced by cyclic differentials 
(as in cyclic velocities with $\d \lambda$'s deleted from their denominators).  
One then also has to consider the constraints in `Jacobian' ($Q^{\sfC}$, $\d Q^{\sfC}$) variables rather than Lagrangian ones.

Configurational Relationalism usually interferes with the Frozen Formalism Problem since it needs to be tackled prior to it for the above resolution's Machian emergent time 
to be free of reference to the $G$-auxiliaries.
This is manifest in the next formula below.  
Since it is in this context that Barbour--Bertotti's work \cite{BB82} is further necessary, I denote this timefunction by $t^{\se\sm(\sJ\sB\sB)}$. 
E.g. for GR, $G$ = Diff($\Sigma$) -- the 3-diffeomorphisms of the constant-topology spatial manifold -- as encoded by the auxiliary `frame variable' $F$ in (\ref{S-Jac}). 
Then $F$ then enters the expression for the emergent time, from where it has to be cancelled out with the following extremization:
\beq
\lt^{\se\sm(\sJ\sB\sB)} = \stackrel{\mbox{\scriptsize extremum}}{\mbox{\scriptsize $F$ $\in$ Diff($\Sigma$) of $S^{\tG\tR}_{\tr\te\tl\ta\ttt\ti\to\tn\ta\tl}$}} 
\left(
\int \left.||\d_F(h, \phi)||_{\mbox{\boldmath\scriptsize $\cal M$}}\right/\sqrt{\mbox{Ric}(x; h] - 2\Lambda - |\pa\phi|^2 - V(\phi)}
\right) \mbox{ } .
\eeq
Thus such expressions are only explicit if one succeeds in solving the linear $G$-constraints for the auxiliary $G$-auxiliaries themselves.

\mbox{ } 

\noindent Let us comment further on the form of this expression.  It says that one has to extremize {\sl one} functional in order to use the extremal value from that in 
a {\sl second} functional. 
This is more complicated than the usual variational problem that involves extremizing a single functional. 
One might wonder what happens if one simply extremizes the time functional itself.  
In simple examples such as relational mechanics, this makes no difference \cite{FileR}, but in the general case it produces an answer that a) different and b) inconsistent 
with the outcome of reordering the reducing and time-forming maps.  
The presently given 2-functional extremization, on the other hand, is compatible with such reorderings of the maps involved \cite{FileR}.  

\mbox{ }  

\noindent Finally, Best Matching generalizes further in the map-ordering sense to Configurational Relationalism. 
By `map-ordering sense', I mean that the reduction can be performed at {\sl any} level (e.g. Hamiltonian or quantum-mechanical) 
rather than just specifically a Lagrangian- or Jacobian-level reduction as just described.   

\mbox{ } 

\noindent On the one   hand, it should be said that the present paper's minisuperspace model straightforwardly avoids Configurational Relationalism by its assumption of homogeneity. 
\noindent On the other hand, the present program's PoT resolving strategy has the virtue of succeeding in dealing with Configurational Relationalism.  
In particular, this facet is both present and resolvable for the more realistic and hence important Halliwell--Hawking model that is the objective of the {\sl next} 
papers of this program.

\subsection{Different uses of the word `relational'}

\noindent The word `relational' is used rather differently by e.g. Barbour, Gryb and I \cite{BB82, B94I, RWR, B11, GrybTh, FileR, ARel} on the one hand, 
and by e.g. Crane and Rovelli \cite{Crane, Rov96, Rovellibook} on the other hand.  
There are further distinctions within each of these two groupings (see Figure \ref{BACR}).  

{            \begin{figure}[ht]
\centering
\includegraphics[width=0.5\textwidth]{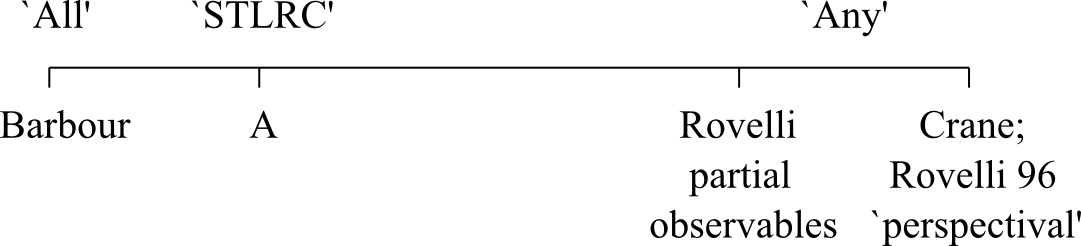}
\caption[Text der im Bilderverzeichnis auftaucht]{\footnotesize{{\bf Types of relationalism}.  This paper concerns the left-hand-side grouping's meaning of `relational'; 
moreover, my specific position (A) also lies somewhere in between Barbour's and Rovelli's, though it is closer to Barbour's.  

\noindent Barbour's and my type of positions are illustrated by the above exposition of Temporal and Configurational Relationalism. 

\noindent Crane's and Rovelli 1996's \cite{Rov96} positions I would term `perspectival', due to their consideration of sets of subsystems rather than just the system as a whole itself.  
This carries connotations (especially at the quantum level) of meaning that what is observed depends on the particular specifics of the observer involved.

\noindent Finally, Rovelli's distinct partial observables position \cite{Rovellibook, Rfqxi} concerns one dynamical variable being employed as the entity with respect to which 
the other dynamical variables evolve.

\noindent Both of the previous sentences' positions concern subsystems in some sense.
This is more strongly so for the first of these (localized subsystems) than for the last of these (subsystem in the sense of 1 degree of freedom out of many, which is
not necessarily a local one, e.g. it could be the scalefactor or a homogeneous matter mode that is employed in this role).  

\noindent The left-hand-side grouping of approaches follow from more long-standing themes in Foundations and Philosophy of Physics.
In contrast, the right-hand-side ones (especially the partial observables position) have so far been more widely used in Quantum Gravity programs.  

\noindent I note furthermore that some parts of the left-hand-side and right-hand-side groupings of positions are mutually compatible and thus can be considered at once \cite{FileR}, 
whilst others require a choice to be made.
The `any change', `all change' and STLRC fork of Sec \ref{TR} is an example of such a choice (in the correspondence indicated at the top of the figure), 
and also one of the ways in which  my position lies in between the other two.
A consequence of these differences in conceptualizing relationalism is that the corresponding timeless formulations of physics split into 
separate families as delineated in \cite{Kuchar92I93, APOTAPOT2}.

\noindent See my review \cite{ARel} for further comparison of these and yet further positions on relationalism.} }
\label{BACR} \end{figure}          }

\subsection{Four further Problem of Time facets}

\noindent A third PoT facet is the Problem of Beables.  
Observables, or beables -- following John Bell \cite{Bell}: a more cosmologically and quantum-mechanically appropriate notion than observables -- 
are hard to come by in classical and quantum GR. 
Models with trivial Configurational Relationalism (or resolved Best Matching) readily possess an explicit full set of classical {\it \K beables}.
I.e. quantities that Poisson-brackets-commute with the classical linear constraints.    
For the present paper, these are 
\beq
\mbox{K = {\cal F}[$\Omega, \phi, p_{\Omega}, p_{\phi}$ alone]} \mbox{ } .  
\eeq
I.e. the \K beables are functionals of the scale variable, the scalar matter field and the conjugate momenta of each.
\noindent In the case of trivial Configurational Relationalism, Halliwell furthermore provided \cite{H03, H09H11} a prescription for Dirac beables 
-- commuting with the quadratic constraint also -- which I promoted to the case of models with resolved Best Matching too \cite{AHall, FileR}.

\mbox{ }  

\noindent The other three facets to overcome in order to have a local resolution of the PoT are as follows. 

\noindent A fourth facet is the Constraint Closure Problem: that the quadratic constraint following from Temporal Relationalism and the linear constraints following from Configurational 
Relationalism are all the constraints.
An alternative possibility is that the Dirac procedure \cite{Dirac} gives back further constraints, including the possibility of inconsistency.   
 
\noindent A fifth facet is the Foliation Dependence Problem: such dependence would be regarded as contrary to GR's capacity to include the perspectives of all observers. 

\noindent A sixth facet is the Spacetime Reconstruction Problem. 
This is from the assumption of just space or from dropping the assumption of the continuum. 
It has further quantum-level motivation \cite{Battelle}, 
from configurations fluctuating quantum-mechanically and deformed geometries generally not all being embeddable within a common spacetime.

\mbox{ }  

\noindent See \cite{FileR} for a more full account of these three facets.  
In any case, at the classical level, these three facets are avoided for full GR.  
This is by three results about the algebraic structure of the constraints (see \cite{AM13} for details and \cite{Dirac, T73, HKT, RWR, Phan-Lan2} for the original references). 
In particular, the Dirac algebroid of the GR constraints closes and Foliation Dependence is resolved by demonstrating Refoliation Invariance [Figure 1d)].  

\mbox{ } 

\noindent However, for minisuperspace in general, there is a simpler triple of results resolution based on homogeneity \cite{ASand}.  
Homogeneity implies no momentum constraint, so the Hamiltonian constraint stands alone, giving closure. 
Homogeneity implies a foliation privileged by the surfaces of homogeneity, and the spacetime to be reconstructed in this case is simpler due to this structure too.
The bracket of two Hamiltonian constraints has a momentum constraint-factor that is zero by homogeneity, thus one has a bona fide algebra of constraints  
(rather than the constraint algebroid of full GR).
Furthermore, the obstruction term to classical spacetime reconstruction \cite{RWR, Phan-Lan2, AM13} contains a spatial derivative factor, so it is also trivial for homogeneous models.  
[The fourth to sixth facets are mostly mentioned due to their nontrivial nature in applying the present program to the Halliwell--Hawking model.]

\subsection{`A local' resolution of the Problem of Time}

\noindent I next explain what I mean in this article by `a local' resolution of the PoT.  
The PoT is taken to have 8 facets. 
`A local' is then a conceptually well-defined subset of six of these, i.e. to the exclusion of the Multiple Choice Problem Facet by use of `a', 
and of the Global PoT Facet by use of `local'.

\mbox{ } 

\noindent The Multiple Choice Problem is a purely quantum-mechanical problem following from classical equivalence under canonical transformations in general not implying quantum-mechanical 
equivalence under unitary transformations.  
[I take `Multiple Choice Problem' in the specific sense of Kucha\v{r} and Isham \cite{Kuchar92I93}; this means from a technical perspective that it is chiefly based on the 
Groenewold--Van Hove phenomenon \cite{Gotay}.]   
Consequently a single classical theory can lead to multiple inequivalent quantum theories. 
This is a time issue through different classical choices of time variables being capable of leading to quantum inequivalence, and also via this problem occurring in one's choice of 
beables/observables.

\mbox{ }

\noindent The Global PoT itself has many subfacets. 
It can refer to both non-globality over space or in one's notion of time itself. 
Also it can refer to the difference between infinitesimal and extended regions, or to topological effects.  
In the PoT context, whichever of these types of globality can apply to timefunctions, frames, foliations, transformations, validity of solutions of equations, beables/observables and more.
A detailed treatment of global issues and their classification would be an article-length account itself, and is indeed one I am preparing to write an article on in 2014.  
A few pertinent comments as regards this paper's model are that the action and the emergent time notions are in general only intended to hold locally, as are the h-l split 
and the notions of beables employed.  
Some of these issues can be furtherly subjected to `patching together' of regions (a concept more fully laid out in e.g. \cite{Bojowald}\footnote{Whilst \cite{Bojowald} 
also consider a `Multiple Choice Problem', these references do not make any mention of the Groenewold--Van Hove phenomenon.
Thus `Multiple Choice Problem' in \cite{Bojowald} is not meant in the specific sense of the Kucha\v{r} and Isham notion of `Multiple Choice Problem' that I follow in the present program.} 
).
Finally, note that this use of `local' in the present paper is not to be confused with `local degrees of freedom' in the field-theoretic sense, since in particular 
minisuperspace models do not possess any such.

\section{Machian Semiclassical Approach}

\subsection{Quantum version of our minisuperspace model}

\noindent Kinematical quantization for this system involves what is to be expected for a system whose isometry group is the 2-$d$ Poincar\'{e} group.  
I.e. a boost $K$ and a 2-vector $P$ = ($p_{\Omega}, p_{\phi}$) that form this isometry group and another 2-vector $X$ upon which this group acts.  
This is the indefinite counterpart of the familiar Heisenberg group (here 2-$d$).  

\mbox{ } 

\noindent Dynamical Quantization is based on the relationally-motivated \cite{Banal} conformal ordering \cite{Magic}.  
Moreover, either the 2-$d$ness or the flatness of the configuration space suffices to ensure that the conformal ordering is equal to the Laplacian ordering. 
Thus the wave equation is the toy model Wheeler--DeWitt equation of the form 
\beq
\hbar^2 \Box_2\Psi := \hbar^2\triangle_{\mathbb{M}^2}\Psi = \hbar^2\{\pa_{\Omega}^2 - \pa^2_{\phi}\}\Psi = \mbox{exp}(6\Omega)\{\mbox{exp}(-2\Omega) - 2\Lambda - V(\phi) \}\Psi \mbox{ } .
\eeq
\noindent This is elementary to solve for $V(\phi) = 0$, and then one can treat exp$(6\Omega)V(\phi)$ as an interaction term to which one can apply a standard form of 
time-independent perturbation theory.  

\noindent Moreover, Frozen Formalism Problem resolutions by $t^{\se\sm(\sJ)}$ fails at the quantum level since the quantum wave equation remains in the form $\widehat{\cal H}\Psi = 0$. 
\noindent My answer to that is to Machianize semiclassical approach.  
The timestandard this produces is indeed similar to $t^{\se\sm(\sJ)}$ from a Machian perspective. 
Furthermore these two timestandards are indeed expected to differ on Machian grounds also, as explained in the next SSec.

\subsection{The standard semiclassical approach}

\noindent Here \cite{HallHaw, Kieferbook} one again has a h--l split. 
Moreover now the h carry additional connotation of `slow' to l's `fast' due to the double nature of the wavefunction ansatz and associated approximations in use. 
Namely, one now has the {\it Born--Oppenheimer (BO)--WKB ans\"{a}tze}
\beq
\Psi(\mh, \ml) = \psi(\mh)|\chi(\mh, \ml)\rangle \mbox{ } , \mbox{ } \mbox{ } \psi(\mh) = \mbox{exp}(i\,W(\mh)/\hbar) \mbox{ }  . 
\label{2-ans}
\eeq 
\noindent One then forms the h-{\it equation},  
\beq
0 = \langle\chi| \widehat{\cal H} \, \Psi = -\{\pa_{\Omega}W\}^2 + 2i\hbar\pa_{\Omega}W\langle\pa_{\Omega}\rangle + \hbar^2\langle\Box_2\rangle + 
\mbox{exp}(6\Omega)\{\langle V(\phi)\rangle + 2\Lambda - \mbox{exp}(-2\Omega)\}  \mbox{ } .
\label{h-W}
\eeq 
(applying the product rule and the WKB approximation).  
Then interpreting $W$ as a Hamilton's function and applying the momentum--velocity relation, this is cast in the form
\beq
-\mbox{exp}(6\Omega)\Omega^{*\,2} + 2i\hbar\,\mbox{exp}(3\Omega)\Omega^* \langle\pa_{\Omega}\rangle + \hbar^2\langle\Box_2\rangle + 
\mbox{exp}(6\Omega)\{\langle V(\phi) + 2\Lambda - \mbox{exp}(-2\Omega)\} = 0  \mbox{ } .
\label{h-E}
\eeq
for $* := \pa/\pa t^{\se\sm(\sW\sK\sB)}$. 
Under a number of simplifications, the most common approximand to (\ref{h-W}) is a Hamilton--Jacobi equation and to (\ref{h-E}) an energy equation. 
Integrating the latter gives a zeroth-order expression for the emergent WKB time $t^{\se\sm(\sW\sK\sB)}_{(0)}$ that clearly coincides to this order 
with $t^{\se\sm(\sJ)}$, and thus also fails to be Machian. 
The semiclassical correction terms, however, are distinct from the classical ones; both are Machian and the difference between them is for Machian reasons too: now it is
{\sl quantum} change that needs to be given the opportunity to contribute.  
Thus the semiclassical and classical Machian timestandards must indeed be distinct \cite{FileR, ACos2}.

Before treating these correction terms, however, it is useful to additionally form the l-{\it equation} 
\beq
0 = \{1 - |\chi\rangle\langle\chi|\}\widehat{\cal H}\,\Psi = \{1 - |\chi\rangle\langle\chi|\}\{2i\hbar \, \pa_{\Omega}W \, \pa_{\Omega} + \hbar^2\Box_2 + \mbox{exp}(6\Omega)\} |\chi\rangle \mbox{ } . 
\label{fluc-eq}
\eeq 
Thus this originally comes in the form of a fluctuation equation. 
However, it can be recast, modulo further approximations, and the chroniferous move\footnote{Here $M_{\th\th}$ is  
the $\sh\sh$ component of the configuration space metric with inverse $N^{\th\th}$.}
\beq
N^{\sh\sh}  i\hbar  \frac{\pa W}{\pa \mh}               \frac{\pa \left| \chi\right \rangle}{\pa \mh} = 
i\hbar \, N^{\sh\sh}  p_{\sh}                           \frac{\pa \left| \chi\right \rangle}{\pa \mh} =
i\hbar \, N^{\sh\sh}  M_{\sh\sh}  \Last \mh             \frac{\pa \left| \chi\right \rangle}{\pa \mh} = 
i\hbar \frac{  \pa \mh  }{ \pa t^{\se\sm(\sW\sK\sB)} }  \frac{\pa \left| \chi\right \rangle}{\pa \mh} = 
i\hbar                                                  \frac{\pa \left| \chi\right\rangle }{\pa t^{\se\sm(\sW\sK\sB)}}    \mbox{ } , 
\eeq
into an emergent-WKB-TDSE (time-dependent Schr\"{o}dinger equation) for the l-degrees of freedom   
\beq
i\hbar  \, \mbox{exp}(3\Omega)   \frac{\pa|\chi\rangle}{\pa t^{\se\sm(\sW\sK\sB)}} = \widehat{H}_{\sll} |\chi\rangle    \approx  
\frac{\hbar^2}{2}  \pa_{\phi}^2   |\chi\rangle    +     \frac{\mbox{exp}(6\Omega)}{2}  V(\phi) |\chi\rangle  \mbox{ }
\label{TDSE2} \mbox{ } . 
\eeq
The emergent-time-dependent left-hand side of this arises from the cross-term $\pa_{\sh}|\chi\rangle\pa_{\sh}\psi$. 
$\widehat{H}_{\sll}$ is the remaining surviving piece of the quadratic constraint that acts as a Hamiltonian for the l-subsystem.  

\noindent (\ref{TDSE2}) is, modulo the h--l coupling term, `ordinary relational l-physics'.  
The more usually considered simpler situation `has the scene set' by the h-subsystem for the l-subsystem to have dynamics. 
This dynamics is furthermore slightly perturbed by the h-subsystem, while neglecting the back-reaction of the l-subsystem on the h-subsystem.  
One might even argue for the interaction term to be quantitatively negligible as regards the observed l-physics. 
\noindent Thus, overall, the fluctuation l-equation (\ref{fluc-eq}) can be rearranged to obtain a TDSE with respect to an emergent time that is `provided by the h-subsystem'.

\subsection{Comparison to previous WKB and/or BO split minisuperspace models}

\noindent In all of below, scale is h and matter fields are l.
Such models started with simple minimally-coupled scalar field matter in the works of Banks \cite{Banks} and of Padmanabhan and Singh \cite{PS89}.
Brout and Venturi \cite{BV89} specifically used a BO approximation as well as a WKB one, and formed and kept quantum-cosmological Berry phase terms. 
Back-reaction terms involving higher derivatives were kept in Kiefer and Singh \cite{KS91}, which work was more recently followed up by Kiefer and Kramer \cite{KK}; 
back-reaction terms were also considered in \cite{Casadio-98}.  
Diabatic terms were kept instead in work of Massar and Parentani \cite{Parentani}. 
Non-minimally coupled scalar fields were considered by the Bologna Group \cite{ACG99}, who followed this up by considering
a minimally-coupled scalar inflaton alongside a conformally-coupled scalar field \cite{AACST07}.  
The above scalar field models were briefly reviewed in Zeh's book \cite{Zehbook} and more extensively in Kiefer's \cite{Kieferbook}; 
see also the recent \cite{KTV13} for BO-approximated Quantum Gravity corrections to inflationary cosmology.

\mbox{ } 

\noindent Ultimately, however, in Cosmology one is more interested in modelling small anisotropies and even more so in modelling small inhomogeneities. 
Semiclassical quantum cosmology with small anisotropies in the l-role were considered in vacuo by e.g. Marolf \cite{Marolf94}; \cite{A-MSS-2} will consider these in 
the presence of a scalar field also (now treated as h jointly with the scalefactor). 
Semiclassical quantum cosmology with small inhomogeneities is well-known to have started with Halliwell and Hawking's paper \cite{HallHaw}; 
this was followed up by Halliwell's paper on correlations \cite{Halliwell87} and Kiefer's on decoherence \cite{Kiefer87}.  

\mbox{ } 

\noindent As regards emergent semiclassical time, this was first widely brought to attention by Banks \cite{Banks} and Halliwell--Hawking \cite{HallHaw}, 
through it can be traced back to DeWitt's much earlier work \cite{DeWitt67}.
This SSec's models do not however present small Machian corrections to the cosmic time; 
these are novel to the present paper's program, as is the use of rectified time in the next SSec.
On the other hand, \cite{Halliwell87}'s correlations exemplify a timeless approach.
Via decoherence in some such approaches \cite{Kiefer87, Giu, Kieferbook} one instead can take a histories-theoretic (see Sec 5) attitude to the issue of time in 
Quantum Cosmology.
Finally, the above three perspectives on time can be combined \cite{H03, H09H11} (and Sec 5); 
moreover this combination can continue to be considered from a Machian perspective \cite{AHall, FileR, CapeTown12}.

\subsection{Rectified Time}

\noindent Next, note that a rectified time given by
\beq
\mbox{exp}(3\Omega)\pa/\pa t^{\se\sm(\sW\sK\sB)} = \pa/\pa t^{\se\sm(\sr\se\scc)}
\eeq
simplifies the l-equation to the {\it emergent rectified-TDSE} 
\beq
i\hbar \, \frac{\pa|\chi\rangle}{\pa t^{\se\sm(\sr\se\scc)}} \approx  
\frac{\hbar^2}{2}\pa_{\phi}^2  |\chi\rangle + \frac{\mbox{exp}(6\Omega)}{2}V(\phi) |\chi\rangle  \mbox{ } .  
\eeq
Moreover, because the h and l equations for a coupled set of equations that in general have to be treated together 
as a system to solve for the unknowns $t^{\se\sm}(\sr\se\scc)$ and $|\chi\rangle$, I then choose to write the h-equation in terms of the rectified time as well, 
\beq
- \{\d\Omega/\d t^{\se\sm(\sr\se\scc)}\}^2 + 2i\hbar\{\d\Omega/\d t^{\se\sm(\sr\se\scc)}\} \langle\pa_{\Omega}\rangle + \hbar^2\langle\Box_2\rangle + 
\mbox{exp}(6\Omega)\{\langle V(\phi)\rangle + 2\Lambda - \mbox{exp}(-2\Omega)\} = 0 \mbox{ } . 
\eeq
Only now do I integrate the h-equation, to obtain in general 
\beq
\lt^{\se\sm(\sr\se\scc)} = \int\d\Omega\big/\big\{i\hbar\langle\pa_{\Omega}\rangle + \sqrt{\hbar^2\{\langle \Box_2 \rangle - \langle\pa_{\Omega}\rangle^2\} 
+ \mbox{exp}(6\Omega)\{\langle V(\phi) \rangle + 2\Lambda - \mbox{exp}(-2\Omega)\}}\big\}  \mbox{ } .  
\eeq

\subsection{Machian emergent semiclassical time}

\noindent The zeroth approximation here coincides with the classical zeroth approximation, which was already declared to be non-Machian.
\noindent I then use the binomial expansion and expansion in powers of $\hbar$ and of the number of QM averaged/expectation terms 
to isolate what will often serve as first correction terms: 
\beq
\lt^{\se\sm(\sr\se\scc)}  \approx \lt^{\se\sm(\sr\se\scc)}_{(0)} 
- \frac{\langle V(\phi) \rangle}{2}\int\frac{\d\Omega}{\mbox{exp}(3\Omega)\{2\Lambda - \mbox{exp}(-2\Omega)\}^{3/2}} 
- i\hbar\int\frac{\d\Omega \langle\pa_{\Omega}\rangle}{\mbox{exp}(6\Omega)\{2\Lambda - \mbox{exp}(-2\Omega)\}}          + O(\hbar^2, \langle\mbox{ }\rangle^2)\mbox{ } . 
\eeq
These are new to this paper in the case of minisuperspace.
The first of these contains an integral factor that can be evaluated without knowledge of the l-subsystem.  
This gives explicitly 
\beq
\frac{-1}{\sqrt{2\Lambda\mbox{exp}(2\Omega) - 1}} + \mbox{arctan}
\left(
\frac{1}{\sqrt{2\Lambda\mbox{exp}(2\Omega) - 1}}
\right) 
\mbox{ } .  
\eeq
For computation of its co-factor, or the other term, we need to address the l-subsystem too. 
\noindent I then consider 1) the free approximation to the emergent rectified-TDSE and 2) emergent rectified-time-dependent perturbations about this. 
Note that for terms with already other small factors, the perturbations only contribute to lower order, so we can omit that part of the calculation for now.  
\noindent 1) involves calculating the expectation of $V(\phi)$ can be considered by Fourier expansion and then 3-trig-function integrals.
\noindent 2) can be evaluated by the method given in \cite{QuadIII}, giving
\beq
\frac{1}{3}\frac{\mbox{exp}(-4\Omega) + 8\Lambda \mbox{exp}(-2\Omega) - 32 \Lambda^2}{\sqrt{2\Lambda - \mbox{exp}(-2\Omega)}} \mbox{ } .
\eeq  
\noindent See \cite{SemiclIII, FileR, ACos2, A-MSS-2} for more details of the physical justification of, and mathematical methods for, 
some of the next most simple regimes for such a semiclassical scheme. 
\cite{A-MSS-2} in particular does so for this paper's minisuperspace example and its Bianchi IX counterpart.  

\mbox{ }

\noindent The present paper's approach involves a complex notion of time. 
Another approach that has recently given a complex notion of time is Bojowald et al's approach \cite{Bojowald}. 
There is therefore some common interest in what a complex notion of time means conceptually, though in greater detail these two approaches are distinct.  
Complex time can carry connotations of nonunitarity \cite{Kieferbook}. 
\cite{Bojowald}'s complex time is related to that approach's choice of inner product and time variable; this also happens in the Klein-Gordon counterpart of this approach 
so it is not necessarily related to non-unitarity. 
However, in this case complex time does show up in tandem with non-unitarity of the dynamics setting in. 
On the other hand, in my approach to semiclassical models the complex time is tied to the approximate nature of the wave equations considered.  
This arises from the semiclassical h-equation being a quantum-corrected version of the quadratic constraint.  
Then since quantum equations are in general complex, rearranging one to form a candidate timefunction is capable of producing a timefunction with complex corrections.  
More generally in semiclassical approaches, the presence of non-unitarity is not related to the semiclassical approximation itself, 
but rather to the choice of the time variable \cite{H86}.

\subsection{WKB problem}

\noindent N.B. that the working leading to such a TDWE (time-dependent wave equation) ceases to work in the absence of making the WKB ansatz and approximation. 
Additionally, in Quantum Cosmology, this is not known to be a particularly strongly supported ansatz and approximation to make \cite{Zeh, Zehbook, Kuchar92I93, APOTAPOT2}.    
This is crucial as regards the present Article and program, since propping this up requires considering one or two further PoT strategies from the classical level upwards.  

\noindent Why is it not strongly supported in Quantum Cosmology?
Firstly discount the applicability of the circumstances under which a WKB ansatz is used in ordinary QM.  
1) In ordinary QM, one often assumes \cite{LLQM} the WKB ansatz as a consequence of the pre-existence of a surrounding 
classical large system , which is no longer applicable for the whole universe.
2) In ordinary QM, the WKB ansatz can be justified by the lab set-up being a ``pure incoming wave".
But if one assumes a pure wave in the quantum cosmological context, its wavefronts orthogonally pick out a direction which serves as timefunction. 
Thus this amounts to `supposing time' rather than a `bona fide emergence of time' as required to resolve the PoT.

Secondly, the WKB ansatz is not general or a priori natural.  
The $W$-function arises from solving an h-equation that is (at least approximately)  a Hamilton--Jacobi equation. 
However, Hamilton--Jacobi equations have 2 solutions $W^{\pm}$.
Thus in general one would expect not $\phi(\mh)$ = exp(i$W(\mh)/\hbar$) but \cite{Zeh, Zehbook, B93, Kuchar92I93, Giu} a superposition 
$
\phi(h) = A_{+}\mbox{exp}(i\, W_+(\mh)/\hbar) + A_{-}\mbox{exp}({i\, W_-}(\mh)/\hbar)  \mbox{ } .
$
Moreover, the semiclassical approach's trick by which the chroniferous cross-term becomes the time-derivative part of a TDSE is exclusive to wavefunctions obeying the WKB ansatz 
\cite{Zeh86, Zeh, Zehbook, Kuchar92I93, B93, B94II, EOT}. 
Possible ways out of this problem involve decoherence \cite{Giu}, perhaps from histories decohering, causing exp(i$W(\mh)/\hbar$) 
to emerge rather than the above more general superposition.   

Thirdly, the sometimes also mentioned issue of whether exp(i$W(\mh)/\hbar$) is an appropriate solution to a real wave equation in place of a complex one is rendered less severe by the 
simple observation that real equations do however in general admit complex solutions.
This does not however cover the deeper aspect unveiled by Barbour \cite{B93}: 
i.e. it is strange that a real wave equation which does not mix real and imaginary terms leads to a complex equation which does mix these.

\subsection{Handling the rest of a local PoT resolution at the semiclassical level}

\noindent There is no Configurational Relationalism in the first place.  
Quantum \K beables are trivial in the sense of being a subalgebra of the classical \K beables, but nontrivial as regards precisely what this choice entails. 
\noindent Constraint Closure, Foliation-Dependence and Spacetime Reconstruction remain protected at the quantum level by the model's homogeneity.

\section{Combined Machian--Semiclassical--Histories--Records scheme}\label{Combi}

The idea is to get round the WKB problem via combination with the histories and records approaches to the PoT.  
For the present paper's model, records approaches need none of the innovations of \cite{ARec, FileR} since minisuperspace fails to allow for localized structure. 
Histories approaches require but a few departures from the perspective offered by Anastopoulos and Savvidou \cite{AS05}, as follows.
Their approach is a {\sl histories projection operator (HPO)} approach \cite{IL}, applied indeed to the specific case of isotropic minisuperspace models with scalar field matter.  
To make contact with the present paper's work, one needs to specialize their work from FLRW with general spatial curvature to FLRW with positive spatial curvature 
and the usual compensating-sign $\Lambda$-term.  
One also needs to suppress their use of the lapse and its conjugate momentum in order to conform to our approach's Temporal Relationalism.   
These changes do not elsewise affect the classical and quantum histories brackets they consider.
In general, the HPO kinematical quantization is notably different from Sec 3's since i) the HPO approach involves continuous time \cite{AS05}. 
ii) One then gets a 1-$d$ QFT in the time direction.   
Since this is relatively unfamiliar, I clarify it by saying that this has a nontrivial commutation relation not like the $[x_i, p_j] = i\hbar \delta_{ij}$ of finite QM 
but rather of the form $[x_{t_1}, p_{t_2}] = i\hbar\delta(t_1 - t_2)$. 
I.e. a continuous Dirac delta in place of a discrete Kronecker delta, though moreover the former is better replaced by a smeared formulation. 
Following Isham and Linden \cite{IL}, this involves the $L^2_{\mathbb{R}}(\mathbb{R})$ function space of real square-integrable functions on $\mathbb{R}$ in the role of the 
test functions.
Then the nontrivial commutation relation becomes $[x_f, p_g] = i \hbar\int_{-\infty}^{+\infty}f(x)g(x)dx$ for $f, g \, \in \, L^2_{\mathbb{R}}(\mathbb{R})$. 
The ensuing quantum theory is then indeed based on a Fock space as is habitual in QFT; see \cite{IL} for more details.
One can then additionally comply with Temporal Relationalism by formulating one's smearing functions to be parametrization-irrelevant in close parallel to \cite{AM13}.  
(I.e. let the smearing role conventionally played by arbitrary test functions $f, g$ be taken over by new smearing functions that are written in the Temporal Relationalism compatible 
form $\d \xi, \d \zeta$).  

\mbox{ }  

\noindent The combined Machian--Semiclassical--Histories--Records scheme is a particularly interesting prospect \cite{H03} along the following lines 
(see \cite{GMH, H99, H09H11, FileR} for more details).   

\noindent 1) Histories decohereing is a leading (if not as yet fully established) way by which a semiclassical regime's WKB approximation could be legitimately obtained in the first 
place.  
Thus Histories Theory could support the Semiclassical Approach by freeing it of a major weakness.

\noindent 2) There is a Records Theory within Histories Theory, as pointed out by Gell-Mann and Hartle \cite{GMH} and extended to imperfect cases by Halliwell \cite{H99, H03, H09H11}.     
Thus Histories Theory supports Records Theory by providing guidance as to the form a working Records Theory would take.
This also allows for these two to be jointly cast as a mathematically-coherent package.

\noindent 3) Moreover, at the quantum level, in the words of Gell-Mann and Hartle \cite{GMH}, 
\beq
\mbox{Records are ``{\it somewhere in the universe where information is stored when histories decohere}"}. 
\label{situ2}
\eeq
The elusive question of which degrees of freedom decohere which should then be answerable through where in the universe the information is actually stored, 
i.e. where the records thus formed are \cite{GMH, H03}.  
In this way, Records Theory could in turn support Histories Theory. 

\noindent 4) Both histories and timeless approaches lie on the common ground of atemporal logic structures \cite{IL, ID, FileR}.   

\noindent 5) By providing an underlying dynamics or history, the Semiclassical Approach and/or Histories Theory \cite{H03, Kieferbook} overcome present-day pure 
Records Theory's principal weakness of needing to find a practicable construction of a semblance of dynamics or history.
Such would go a long way towards Records Theory being complete.  
Emergent semiclassical time amounts to an approximate semiclassical recovery of the emergent classical time \cite{B94I}, 
which is an encouraging result as regards making such a Semiclassical--Timeless Records combination. 

\noindent 6) The Semiclassical approach aids in the computation of timeless probabilities of histories entering given configuration space regions (see below for more).

\noindent 7) New to the present program, the Semiclassical Approach provides a Machian scheme for Classical and Quantum Records and Histories to reside within.  

\mbox{ }  

\noindent Note 1) At the classical level, inter-relations 1) and 3) are absent since they concern the purely quantum notion of decoherence, 
and 6) vanishes since it concerns a purely quantum probability computation.  
4) is much more trivial now too (standard logic versus nontrivial topos-theoretic intuitionistic logic \cite{ID}).  

\noindent Note 2) I end by I logically ordering the interprotections between the three strategies.
{\sl Meaningless label} histories come first, this gives the Semiclassical Approach and then this gives the emergent-time version of the histories approach.  
Then localized timeless approaches sit inside the last two of these.  
The Semiclassical Approach sits inside the global timeless approach. 
But the global timeless approach can be taken to sit within global meaningless label time histories approach, so down both strands this is primary.  

\mbox{ }

\noindent The current paper's model is archetypal of Halliwell's own examples \cite{H03, H09H11} of this combination, albeit additionally with a Machian interpretation pinned upon it.  
This is by specifically using whichever suitable type of emergent time in the construction of the semiclassical expressions for the timeless probabilities.  
Moreover, this paper's model's case of this construction has no need of the curved-geometry subtleties\footnote{The simplest models to manifest these are the relational 
quadrilateral and the non-diagonal Bianchi IX minisuperspace.} 
indicated in \cite{AHall} since this model possesses a flat presentation of configuration space. 
Being indefinite, this is qualitatively distinguished from Halliwell's absolute particle mechanics examples \cite{H03} and my flat configuration space RPM examples 
\cite{AHall, FileR, CapeTown12}.
Also, possessing no linear constraints, this model has no need to build and subsequently use nontrivial \K beables (in this sense, the triangleland model in 
\cite{AHall, FileR, CapeTown12} is more advanced.   
All in all, one has expressions for semiclassical quantum probabilities such as\footnote{Here $R$ a region of configuration space $\fQ$ with corresponding characteristic function $f_R$.
%
$\bn$ is the normal in configuration space. 
$P$ is a prefactor function, the detailed form of which is given in \cite{HT}.
$\theta$ is a step function and $\epsilon$ is a small number.
0 and f subscripts denote `initial' and `final'.} 
\beq
P_{R}^{\sss\se\sm\si\scc\sll} \approx  
\int \d t^{\se\sm(\sr\se\scc)} \int_{R} \mathbb{D}{\fQ} (\bq) \,\,\, 
\bn^{\sbq} \cdot {\mbox{\boldmath{$\pa$}}} W \, |\chi(\bq)|^2 \mbox{ } 
\label{36}
\eeq
obtained via the Wigner function formulation and various semiclassical approximations. 
$W$ and $\chi$ here are as given in eq (\ref{2-ans}).
For instance, for $R = \{\phi = 0 \mbox{ to } \phi_1 \mbox{ for } \Omega = \Omega_1\}$, the approximately free case $V(\phi) \approx 0$ gives 
\beq
P_{R}^{\sss\se\sm\si\scc\sll} \propto \phi_1 - \frac{\mbox{sin}\, 2\, \mn \, \phi_1}{2\,\mn} \approx \frac{2\,\mn^2\phi_1^3}{3} 
\eeq
with the last approximation holding for the case of $\phi_1$ small.
This approach furthermore provides both classical and semiclassical entities that commute with ${\cal H}$, 
the latter being a subgenre of {\sl class functional}, a simple example of which is 
\beq
{C}_{R} = \theta
 \left(   
\int_{-\infty}^{\infty} \d t^{\se\sm(\sr\se\scc)} f_{R}(\bq(t^{\se\sm(\sr\se\scc)}))  - \epsilon 
\right) P(\bq_{\sf}, \bq_0) \,   
\mbox{exp}(iW(\bq_{\sf}, \bq_0)) \mbox{ } .  
\label{gyr}
\eeq
For a theory with no linear constraints, these suffice as Dirac beables.
For minisuperspace, these can be built in the sense of \cite{H09H11} as well as in the above sense of \cite{H03} 
(the later of which have the advantage of not succumbing to a different kind of frozenness -- the Quantum Zeno Problem \cite{Zeno}).

\section{Conclusion}\label{Concl}

I have shown that a local resolution of the Problem of Time (PoT) that is Machian for the closed spatially-$\mathbb{S}^3$ isotropic minisuperspace GR model with single 
minimally-coupled scalar field matter can be constructed. 
This approach starts at the classical level with an emergent Jacobi time.
This is subsequently corrected at the semiclassical level to form an emergent WKB time.
Next, it is rescaled to an emergent rectified time that further simplifies the semiclassical quantum equations.
Problems with the semiclassical approach on its own are dealt with by passing to a combined Semiclassical--Histories--Records scheme which amounts to placing a Machian interpretation 
on previous works of Anastopoulos and Savvidou \cite{AS05} and of Halliwell \cite{H03}.  

\mbox{ }  

\noindent In contrast with my earlier work on this approach for relational particle mechanics (RPM), the current paper's study is simpler in not having 1) linear constraints. 
2) A nontrivial notion of structure (and thus of structure formation and of localized records).
On the other hand, the present paper's minisuperspace work is aligned with full GR in possessing an indefinite configuration space metric and a potential that is restricted 
from its originating from full GR. 
It is also helpful to provide this minisuperspace example for this Machian approach to the PoT, since minisuperspace is rather more well-known than RPM's. 
This begins to demonstrate the widespread applicability of this approach to the modelling of our universe and of background-independent physical theories more generally. 
The next step in this program is to complete the diagonal Bianchi IX  example \cite{A-MSS-2}, in which the l-role is played by a small amount of anisotropy. 

\mbox{ } 

\noindent The upcoming common focus  \cite{ASand, A-HallHaw-1} of these RPM and minisuperspace works is to the case of the Halliwell--Hawking model \cite{AHall} of inhomogeneous 
perturbations about isotropic minisuperspace.
For this, the present paper's model is the h-part and the new l-part are the inhomogeneous perturbations which may have seeded the galaxies and CMB hot spots that we observe.   
The Halliwell--Hawking model is of inhomogeneous perturbations about homogeneous cosmology. 
It is already a realistic model for the possible quantum-cosmological origin of large-scale structures such as galaxies and CMB hot-spots 
(via amplification by some inflationary mechanism \cite{inflation-13}).
This model possesses nontrivial linear constraints, which, like for RPM's, are algebraically solvable \cite{ASand}. 
It also clearly possesses a nontrivial notion of inhomogeneity/clumping/structure. 
From a PoT perspective, it additionally has nontrivial semiclassical Constraint Closure, Foliation Dependence and Spacetime Reconstruction considerations. 

\mbox{ } 

\noindent Finally, as further detailed in \cite{FileR}, one is yet to face, for whichever model arenas, the Global PoT and the Multiple Choice Problem within this Machian paradigm.  
The detailed mathematical status of the Multiple Choice Problem for the present paper's model remains unknown.

\mbox{ } 

\noindent{\bf Acknowledgements}. I thank those close to me for support.  
Jeremy Butterfield, George Ellis, Jonathan Halliwell, Sophie Kneller, Matteo Lostaglio,Flavio Mercati, David Sloan and the Anonymous Referees for comments and discussions.
John Barrow, Jeremy Butterfield, Marc Lachi$\grave{\me}$ze-Rey, Malcolm MacCallum, Claus Kiefer, Don Page, Reza Tavakol and Juan Valiente--Kroon for help with my career.
This work was carried out whilst I was funded in 2011 and 2012 by a grant from the Foundational Questions Institute (FQXi) Fund, 
a donor-advised fund of the Silicon Valley Community Foundation on the basis of proposal FQXi-RFP3-1101 to the FQXi.  
I thank also Theiss Research and the CNRS for administering this grant.  
Some of this work was typed up whilst receiving hospitality from the Institute of High Energy Physics at Protvino, Moscow.



\end{document}